# Optical microelastography: ultrafast imaging of cell elasticity

Pol Grasland-Mongrain[1], Ali Zorgani[2], Shoma Nakagawa[3], Simon Bernard[1], Lia Gomes Paim[3], Greg Fitzharris[3,4], Stefan Catheline[2,*], Guy Cloutier[1,5,*]

**Elasticity is a fundamental cellular property that is related to the anatomy, functionality and pathological state of cells and tissues. However, current techniques based on cell deformation, acoustic force microscopy or Brillouin scattering are rather slow and do not always accurately represent cell elasticity. Here, we have developed an alternative technique by applying shear wave elastography to the micrometer scale. Elastic waves were mechanically induced in live mammalian oocytes using a vibrating micropipette. These audible frequency waves were observed optically at 200,000 frames per second and tracked with an optical flow algorithm. Whole cell elasticity was then mapped using an elastography method inspired by the seismology field. Using this approach, we show that the elasticity of mouse oocyte is decreased when the oocyte cytoskeleton is disrupted with cytochalasin B. The technique is fast (less than 1 ms for data acquisition), precise (spatial resolution of a few micrometers), able to map internal cell structures, robust, and thus represents a tractable novel option for interrogating biomechanical properties of diverse cell types.**

**SIGNIFICANCE STATEMENT:** **In wave physics and especially, seismology, uncorrelated vibrations could be exploited using "noise correlation" tools to reconstruct images of a medium. By using a high-frequency vibration, a high-speed tracking device and a reconstruction technique based on temporal correlations of travelling waves, we conceptualized an optical microelastography technique to map elasticity of internal cellular structures. This technique in rupture with other methods can provide an elasticity image in less than a millisecond, thus opening the possibility of studying dynamic cellular processes and elucidating new mechano-cellular properties.**

The ability to measure the elasticity of a cell provides information about its anatomy, function and pathological state. For example, cell biomechanical properties are related to the cytoskeletal network arrangement, and water content[1]. The cell membrane can harden or soften to modulate passage of biomolecules[2]. Electrochemical activation can induce rapid contraction and mechanical modulation of cell properties in electrophysiology and neurology. Notably, tumor cells are characterized by a change of elasticity[3] and therapies inducing fibrosis, necrosis and apoptosis are also accompanied by changes in tissue elasticity. Cytoskeleton reorganization is also linked to the activation process of immune cells and critical for effective cell-cell interactions, formation of immunological synapses and migration processes[4]. These are just a few examples that emphasize the importance of cell biomechanics in biology.

[1] Laboratory of Biorheology and Medical Ultrasonics, University of Montreal Hospital Research Center, Montréal, Québec, Canada, H2X 0A9;
[2] *LabTAU, INSERM u1032, University of Lyon, Lyon, F-69003, France;*
[3] Oocyte and Embryo Research Laboratory, University of Montreal Hospital Research Center, Montréal, Québec, Canada, H2X 0A9;
[4] Department of Obstetrics and Gynecology, University of Montreal, Montréal, Québec, Canada, H3T 1J4;
[5] Department of Radiology, Radio-Oncology and Nuclear Medicine, and Institute of Biomedical Engineering, University of Montreal, Montréal, Québec, Canada, H3T 1J4.
*These authors contributed equally to this work.

Many techniques have been proposed to measure cell mechanical properties, especially its elasticity. Most need a very accurate model of the cell characteristics but the chosen model may impact the measurement accuracy. Moreover, current measurements take seconds to hours to perform, during which biological processes can modify the cell elasticity, and they necessitate fixing the cell on a substrate. Variations in elasticity by a factor of two can occur within a few seconds[5].

In this study, we propose a new elasticity measurement technique based on elastic wave propagation. This technique performs local measurement of the speed $c_s$ of a shear wave, a type of elastic wave. The shear modulus $\mu$ (elasticity) is given by $\rho c_s^2$ with $\rho$ the medium density, assuming a purely, linearly elastic medium and negligible preloads. Here, we show that the shear wave elastography technique can perform micrometer-scale measurements, and that it can extract local elasticity on a whole cell. Three main technological challenges needed to be met to achieve this: (1) to develop an efficient way to induce kHz-range high-frequency shear waves in cells; (2) to find a robust method to track these waves; and (3) to extract elasticity from the observed traveling elastic waves.

## RESULTS

We first set out to demonstrate that high-frequency shear waves can be induced in cells. The key components of the experimental setup are as follows (Fig. 1): a cell held by a first micropipette and excited by contact with a second micropipette vibrating at 15 kHz with an amplitude of 20 µm; a 100x microscope magnification to observe the cell; and a 200,000 frames per second camera fixed on the microscope to acquire optical images over time. The 15 kHz vibration represents a compromise between a high frequency stimulation to have a wavelength smaller than the cell size, and a low frequency excitation to reduce wave attenuation, especially in such soft medium. The experiment was applied on spherical mouse oocytes (80 µm diameter) that are well characterized and easy to manipulate *ex-vivo*. A finite element simulation with a 15 kHz vibration occurring on the side of a soft solid was also built to validate the technique.

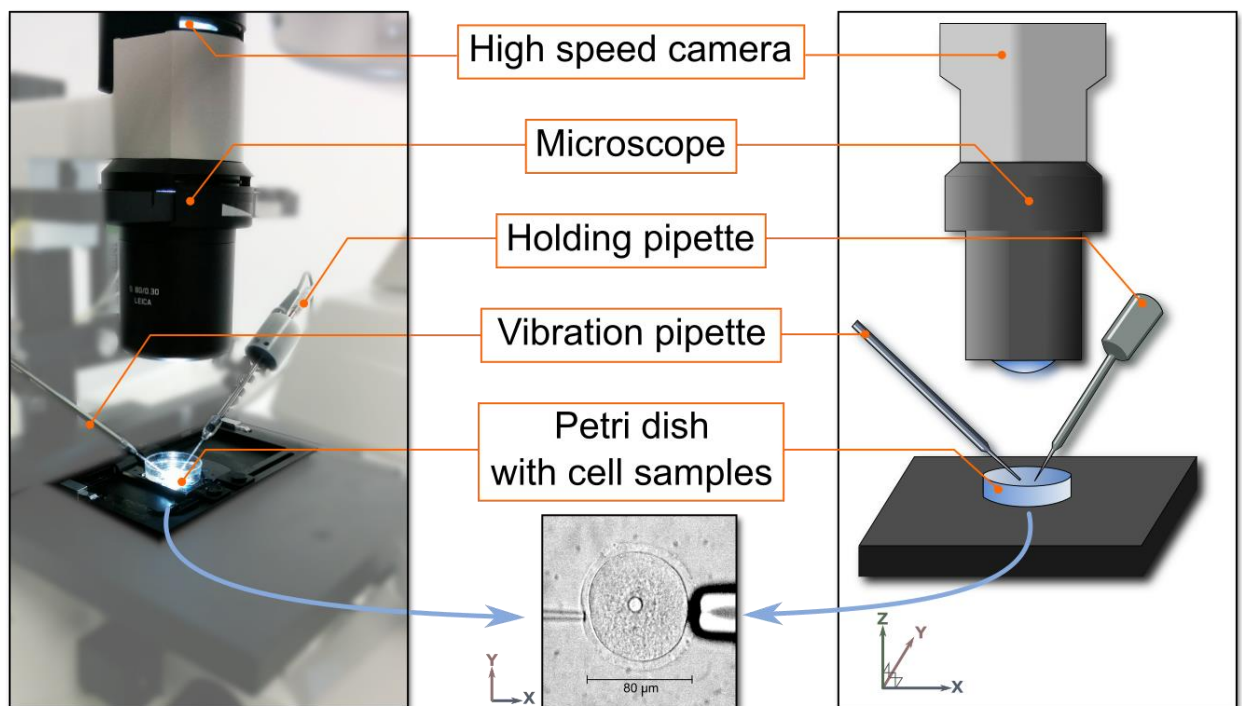

*Fig. 1: Illustration of the experiment (left: picture, right: scheme). A cell placed in a Petri dish is held by a holding pipette (right), and vibration applied using a second pipette attached to a piezo-drive unit (left). Vibration is applied to the zona pellucida of the oocyte. Images of the cell are acquired by a high speed camera through a microscope.*

Using an optical flow algorithm[6], displacements can be seen propagating left to right, with good agreement between experiment (Fig. 2-A) and simulation (Fig. 2-B). Attenuation is strong but displacements can nevertheless be observed on the right side of the cell. Almost no displacement is seen in the surrounding fluid. These displacements propagate at a speed of 1.1 $\pm$ 0.1 m/s, under the form of elastic waves, which could consist in bulk waves (*i.e.*, compression and shear waves) and surface waves (*i.e.*, Rayleigh and Love waves). Any compression waves cannot be seen, as a 15 kHz compression wave has a wavelength of $\approx 10^5$ $\mu$m in such medium, which is thousand times larger than the oocyte diameter. Rayleigh and Love waves also typically propagate at a speed close the shear wave speed (about 10% slower depending on conditions). Consequently, we supposed here that all observed elastic waves propagated at the same speed $c_s$.

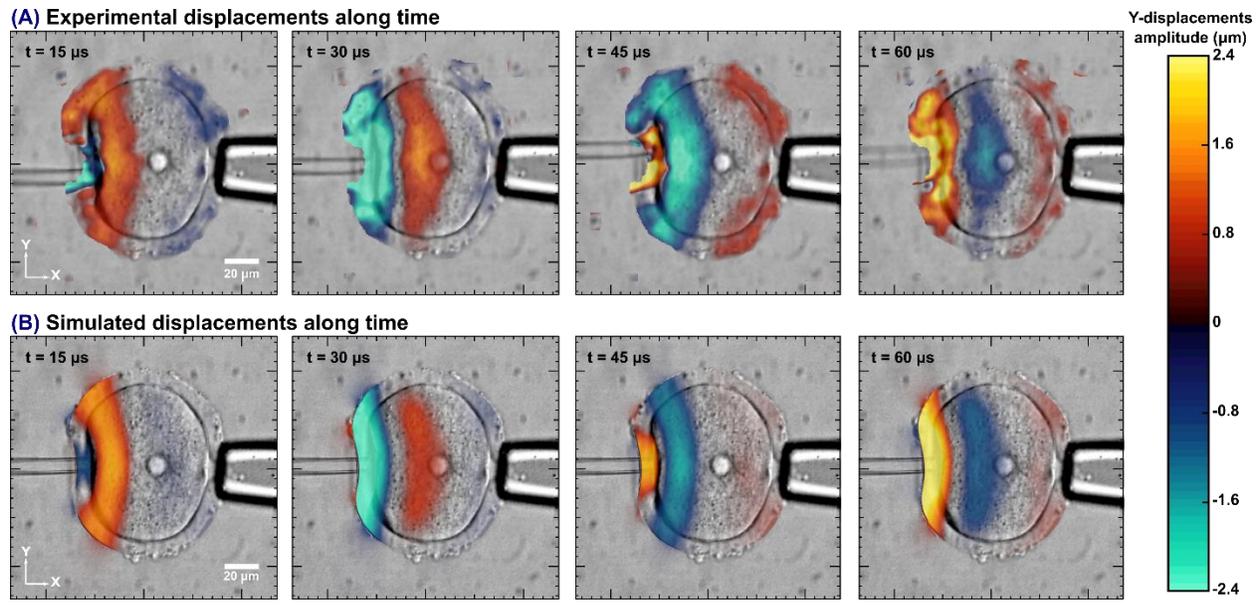

*Fig. 2: Experimental (top) and simulated (bottom) Y-displacement maps, at t = 15, 30, 45 and 60 µs, respectively, superimposed on the optical images of the cell. Displacements with amplitude approximately from -2.4 to 2.4 µm propagating from the left vibrating pipette towards the right side of the cell can be observed. No significant difference of speed is observed between positive and negative displacements.*

To map cell elasticity from observed shear waves, we explored methods proposed in the field of shear wave elastography, such as time-of-flight[7], elastodynamic equation inversion[8,9], and optimal control[10]. In this study, the best reconstruction was obtained using the "passive" elastography algorithm[11,12], inspired by the seismology field - see the Materials and Methods section for details. It primarily calculates the shear wave speed, which allows then estimating the shear elasticity modulus.

On experimental elasticity maps (Fig. 3-A), for analysis purposes, we segmented the oocyte into three functionally distinct zones: the zona pellucida (median of 0.31 kPa), the cytoplasm (median of 0.76 kPa), and the nucleus (median of 0.59 kPa). A median shear modulus was estimated at 0.21 kPa for the extracellular fluid (which should theoretically be zero), but this is attributed to displacements surrounding the cell interpreted by the elastography algorithm as shear waves. Each of these values are pairwise significantly different ($p<0.05$, Mann–Whitney U test). Displacements could however not be properly estimated by the particle imaging velocimetry algorithm near the pipettes due to the high contrast of these objects, so the elasticity couldn't be calculated in a 10-20 um layer around the holding pipette and the actuator.

These elasticity values were used as initial estimates of a simulated medium comprising four concentric circles (Fig. 3-B) representing the different cellular zones. The simulation contained a few artifacts and zones were not homogenous along the X axis; but the algorithm could nevertheless estimate shear moduli with medians close to modeled initial values (differences of 0.04, 0.10, 0.06 and 0.07 kPa, respectively).

To assess reproducibility of the technique, 23 successive measurements were made every 2.5 milliseconds (with 200 images per measurement). Such quick repetition of the measure ensured absence of confounding time-dependent hardening or softening of the cell. Fig. 3-C illustrates the median shear modulus within different zones (cytoplasm, nucleus, zona pellucida, extracellular fluid). No time evolution could be observed in any part of the cell, indicating an excellent reproducibility of the technique.

Next, we applied the elasticity reconstruction algorithm using 5 to 200 frames for each of the 23 measurements to test robustness. Mean elasticity quickly decreased by using 5 to 50 images, then continued to slowly decrease (<10%) with 50 to 160 images, to reach a plateau with around 160 images (Fig. 3-C). This implies that at 200,000 frames per second, reliable measures of cell elasticity require approximately 0.8 ms.

Finally, we studied the impact of the vibration amplitude on the elasticity estimation. With four pipette vibration amplitudes (as measured on optical images), we estimated the shear modulus of one oocyte (Fig. 3-E). The shear modulus of the cell did not depend of the vibration amplitude within the range of 8-20 µm. This result may facilitate future implementation of the technique as it demonstrates that the vibration amplitude is not a critical parameter to consider.

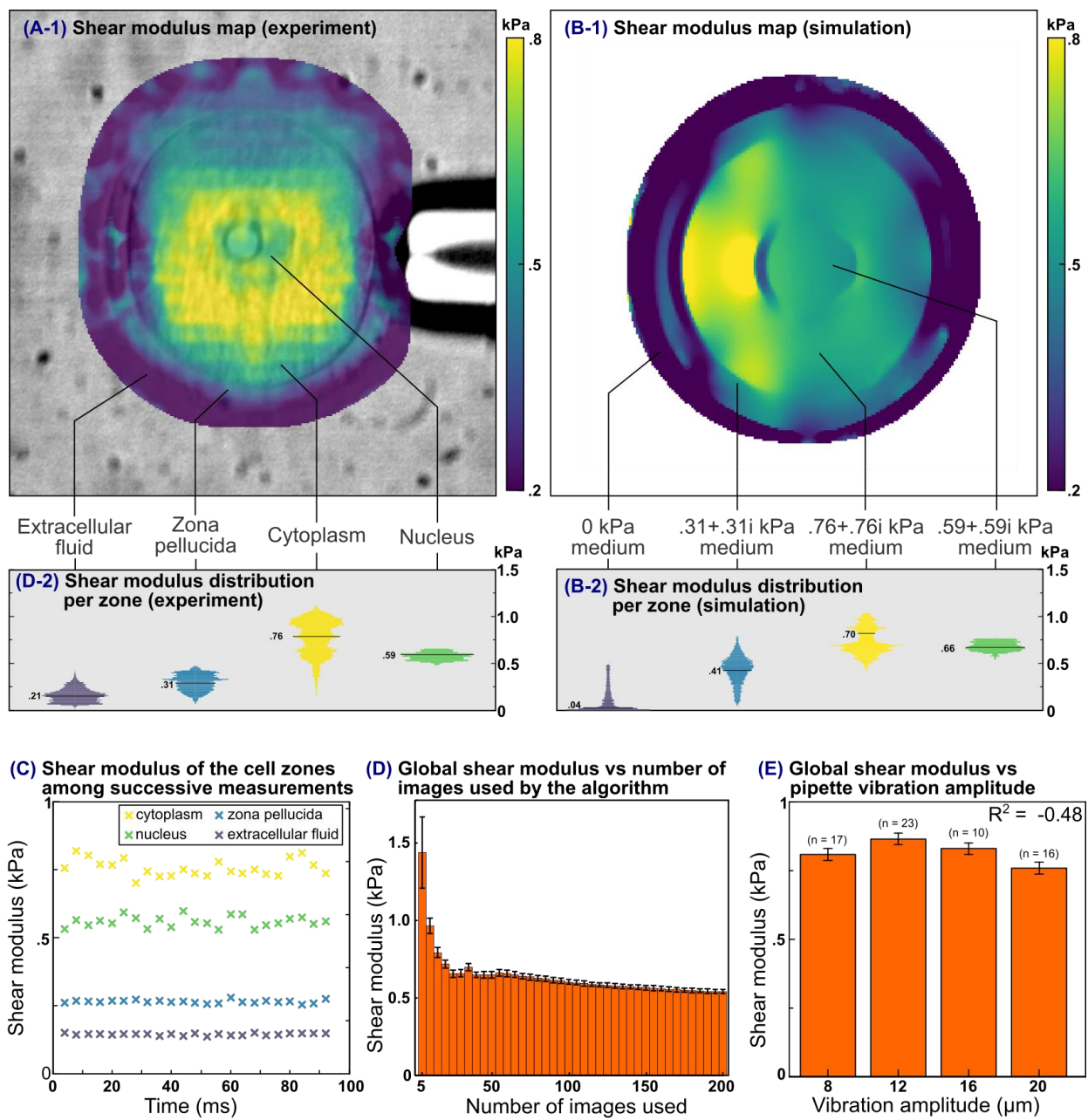

*Fig. 3: (A) Elasticity map estimated from experimental displacements superimposed on the microscopy image (1) and corresponding distribution of elasticity with median values (2) of a mouse oocyte. (B) Elasticity map estimated from simulated displacements superimposed on the microscopy image (1) and corresponding distribution of elasticity with median values (2) of a soft medium mimicking a mouse oocyte. Nucleus, cytoplasm, zona pellucida and extracellular fluid can be easily distinguished on both images; artifacts are observed in the zona pellucida. (C) Median shear moduli for successive measures within different zones (cytoplasm, nucleus, zona pellucida, extracellular fluid). No time evolution is observed. (D) Average median shear moduli of the whole cell among 23 successive measurements as a function of the number of*

*optical images used. Error bars correspond to the standard deviation among successive measurements. (E) Effect of the vibration amplitude on oocyte shear modulus median, obtained by averaging 10 to 23 measures. Error bars correspond to the standard deviation among successive measurements. Pearson's correlation coefficient R is too low to prove any correlation between the measured shear modulus and the vibration amplitude*

Next, we studied the effect of disrupting the actin cortex of the oocyte upon cell elasticity. Cytochalasin B is a toxin known to block polymerization and the elongation of actin, and we thus expected an oocyte softening. Examination of the cortex using alexa-labelled phalloidin and confocal microscopy revealed a major reduction in actin labelling in the oocyte cortex, confirming the expected action of the drug (see supplementary materials). Notably, using optical microelastography, a decrease in shear modulus (softening) was observed when comparing normal and cytochalasin-treated oocytes, both in the cytoplasm and nucleus (Fig. 4-A and 4-B). An artefactual decrease in elasticity within the zona pellucida is observed due to the proximity with the cytoplasm. With a total of 90 measurements on 5 normal cells and 4 cytochalasin-treated cells, we observed a significant decrease ($p<0.002$ with a two-sample *t*-test) of the median shear modulus of the whole cells (Fig. 4-C).

We also observed the shear modulus of cells at different stages of maturation: a germinal vesicle cell (Fig. 4-D), a 2-cell embryo (Fig. 4-E) and a 4-cell embryo (Fig. 4-F). The 4-cell embryo cytoplasm was difficult to segment, as the fourth cell in the background degraded the shear wave displacement estimation. We could nevertheless measure almost the same elasticity at the three stages.

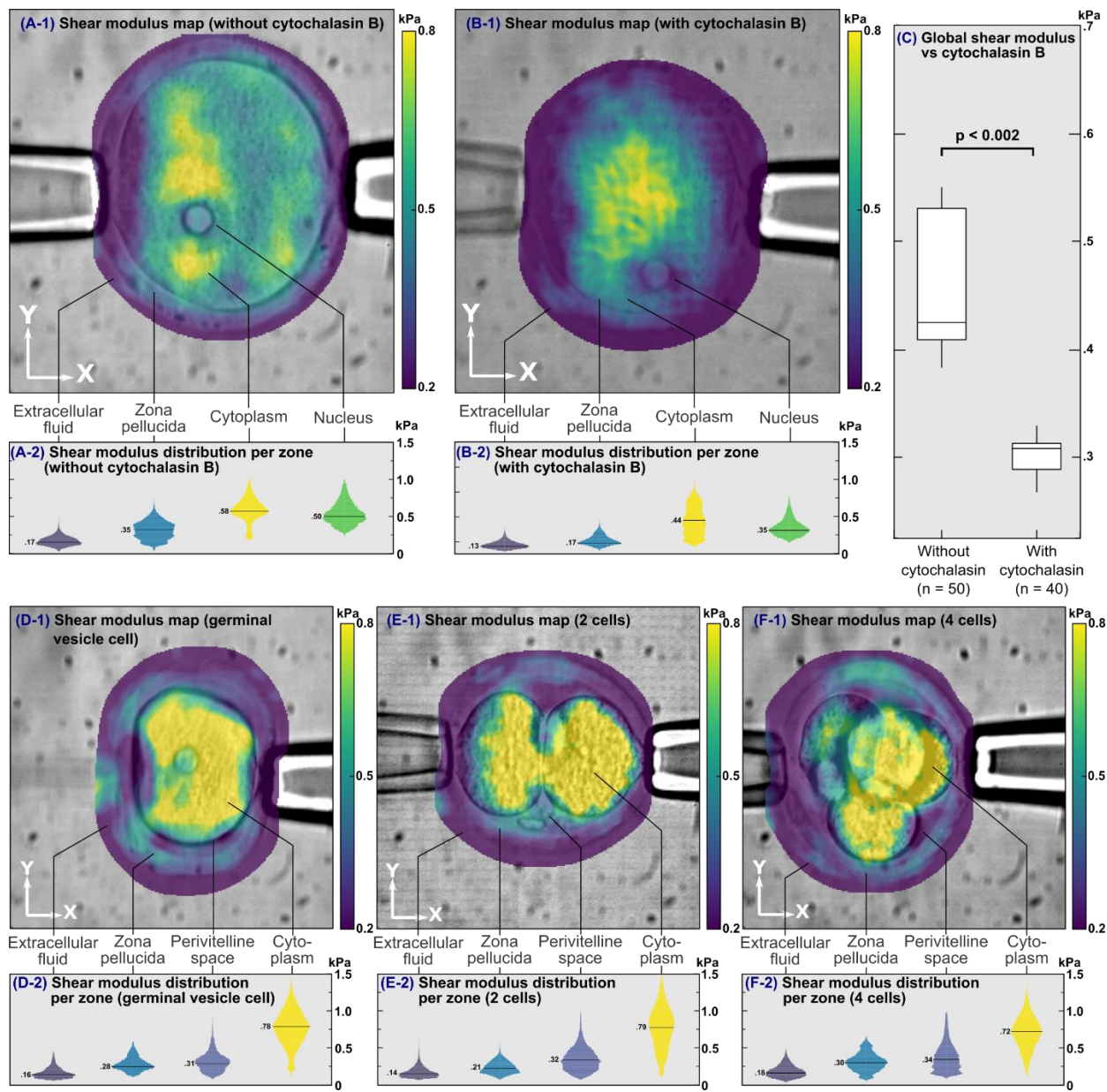

*Fig. 4: Elasticity map superimposed on the microscopy image (1) and corresponding distribution of elasticity with median values (2) of a normal mouse oocyte (A) and a mouse oocyte softened by cytochalasin B (B). Elasticity decreased in all functional areas. (C) Boxplot of whole oocyte shear modulus without and with cytochalasin B, showing a significant decrease in elasticity. Elasticity map superimposed on the microscopy image (1) and corresponding distribution of elasticity with median values (2) of a germinal vesicle cell (D), a 2-cell mouse embryo (E), and a 4-cell mouse embryo (F).*

# DISCUSSION

Compared with existing elasticity mapping techniques, the cellular imaging method presented in this study may become a viable alternative. Indeed, most cell elasticity measurement techniques are based on static cell deformation under an external force, using aspiration micropipettes[13], magnetic bead twisting[14], and optical tweezer or stretcher[15]. The deformation is estimated from optical images acquired before and after applying the force. While global elasticity is easily calculated by dividing the applied force intensity by the measured deformation, localized internal cell structure elasticity is more difficult, because the local deformation depends strongly on the internal distribution of stress. Thus, localized elasticity estimation needs a descriptive model of the cell mechanics, which is difficult to validate and may explain the variability observed among different studies[16-18].

In atomic force microscopy, local elasticity is estimated from the penetration depth of a small probe[19-21]. This technique is able to map elasticity with a sub micrometric spatial resolution. However, measurements are performed at the cell surface only, so that elasticity of internal structures cannot be determined. Besides, elasticity estimation highly depends on the chosen model. In particular, the probe shape must be precisely calibrated, as a change of shape due to probe aging, for example, can have an important impact. In most implementations, acquisitions are rather slow, taking typically at least a few minutes to acquire a full set of data, thus potentially increasing susceptibility to time-dependent confounding biological processes. Moreover, one needs to attach the cell on a substrate to avoid its displacement, thus imposing boundary conditions requiring to be taken into account in the model to avoid biased elasticity measures.

An alternative strategy is to inject fluorescent nanoparticles within the cell, and to measure the mean random displacement of these particles when the cell is subjected to a shear motion[22]. This technique gives access to elasticity and viscosity, but at the location of nanoparticles only, so it does not produce images of the cell viscoelasticity. It also requires a few minutes to make a measurement, and this approach is invasive by nature.

Brillouin scattering microscopy, more recently introduced[23, 24], consists in transmitting a laser beam through the cell, where internal mechanical waves are shifting the laser frequency. Measurement of this shift allows estimating the medium bulk modulus. However, the bulk modulus offers a much smaller contrast than Young's or shear modulus for similar changes in cell elasticity. For example, Scarcelli et al. found a bulk modulus increase of only 16% when the Young's modulus raised by about 500% for the same experimentation[23]. As for abovementioned technologies, acquisitions are rather slow, on the order of the hour.

Contrary to existing methods, the proposed technique does not require a stress distribution model and allows localized absolute elasticity measures (*i.e.*, elasticity in kPa). As it is done in shear wave elastography of whole organ systems, our method may be expended to provide viscous properties[25, 26].

Reported measurements could distinguish internal structures of an oocyte based on their elasticity differences. To our knowledge, no measurement of oocyte elasticity has been previously reported at the frequency we used (15 kHz). Because most biological tissues are exhibiting a frequency-dependent elasticity behavior[27], our measurements lack of gold standard to be compared with. Nevertheless, measured elasticities have been shown to be reproducible and consistent with finite element simulation.

An important advantage of the proposed method is the speed of acquisition, on the order of a few milliseconds. No confounding biological processes, such as cross-linking, could occur during the acquisition time. Future validations of the ultrafast optical microelastography technique may allow demonstrating the capability of this method to follow dynamic cellular processes inducing elasticity changes.

Spatial resolution in shear wave elastography depends on many parameters, such as framerate, reconstruction algorithm, shear wave shape... In the current implementation, variations of elasticity can be observed in the nucleus or in the perivitelline space – so we estimated an average resolution on the order of ten micrometers. Higher spatial resolution, especially to study smaller cells than oocytes, could be obtained by using higher vibration frequency. Different vibrating devices could be conceptualized as multiple harmonic sources would facilitate the elasticity map reconstruction with the proposed method[11, 12]. A three dimensions mapping is also achievable simply by repeating the experiment at different microscope focusing depths.

Finally, the experimental apparatus is rather simple, consisting in a standard microscope, micropipettes and a high speed camera. Phase contrast optical methods might be used for better contrast and resolution. The resolution of the microscope is not a critical parameter: the technique needs mainly to observe and track particles inside the cell. The camera minimum speed of acquisition has to be tuned to the experiment as the shear wave velocity is related to the cell elasticity: we have to observe the shear wave propagation over a few images. A slower less expensive camera could also be employed using a stroboscopic effect, *i.e.*, by repeating synchronized image acquisitions and taking pictures with increasing delays.

Thanks to its speed, robustness and relative simplicity, we therefore envision that this optical microelastography technique could become an alternative for mapping biomechanical properties of cell. It could even open the possibility of studying dynamic cellular processes and elucidating new mechano-cellular properties.

The authors have deposited a patent on the presented technology.

## MATERIALS AND METHODS

### Experimental setup

The ultrafast camera (model v2512, Phantom Research, El Cajon, CA, USA) acquired 256x256 pixel windows at 200,000 images per second. With microscope amplification (Leica DMi1 with 100X lens, Wetzlar, Germany), each square pixel had a lateral size of 0.57 $\mu$m. Cells used were mouse germinal vesicle-stage oocytes. Oocytes were collected as previously[28] and kept in M2 medium (M-7167, Sigma Aldrich, Saint-Louis, MI, USA) supplemented with 200 nM 3-isobotyl-1-methyl-xanthine (I-5879, Sigma Aldrich) during observation. Experiments were performed shortly after oocytes were harvested. Cytochalasin B toxin (C-6762, Sigma-Aldrich) was used to depolymerize the actin cortex in some experiments, as described.

To apply vibration, the oocyte was immobilized using a standard holding pipette. A glass vibration pipette was positioned on the zona pellucida of the oocyte, in an area where the zona pellucida was contacting the plasmalemma. The vibration was created by a piezoelectric device (Piezo impact drive unit, Prim Tech Ltd, Ibaraki, Japan), moving along the Y-axis with a peak-to-peak amplitude of 20 $\mu$m at 15 kHz.

## Finite element simulation

The wave equations were solved using COMSOL (version 3.5a, Stockholm, Sweden) with structural mechanics module, assuming plane strain. We used a 2-dimension model with a manual segmentation of internal structures based on one optical image. Three zones were segmented: the nucleus, a 20 $\mu$m diameter circle; the cytoplasm, an 80 $\mu$m diameter circle around the nucleus; and the zona pellucida, a 10 $\mu$m layer around the cytoplasm. The whole simulated cell was placed in a 200x200 $\mu m^2$ space filled with isotonic saline water. Working area was meshed with approximately 75,000 triangles. We set the bulk modulus for all zones to 2.2 GPa (water compressibility), and the shear modulus elasticity (determined from experimental results) at 0.59 + 0.59i kPa for the nucleus, 0.76 + 0.76i kPa for the cytoplasm, 0.31 + 0.31i kPa for the zona pellucida, and 0 Pa for the surrounding fluid. The nodes on the left side of the zona pellucida were fixed (no displacement in X and Y directions) to simulate the holding pipette, and a vertical prescribed 10 $\mu$m harmonic displacement at 15 kHz was applied to the nodes on the left side of the zona pellucida. The continuity of displacement and strain was ensured by the finite element model code at all other interfaces.

## Elasticity derived from shear wave speed

The proposed technology is inspired by pioneer works in shear wave elastography developed for organ elasticity imaging[7, 8, 29, 30]. Considering a medium as elastic, linear, isotropic and infinite with respect to the wavelength, Navier's equation governs the displacement **u** at each point of the cell:

$$\rho \frac{\partial^2 \mathbf{u}}{\partial t^2} = (K + \frac{4}{3}\mu)\nabla(\nabla \cdot \mathbf{u}) + \mu \nabla \times (\nabla \times \mathbf{u})$$

where $\rho$ is the medium density, **u** the local displacement, $K$ the bulk modulus and $\mu$ the shear modulus. Using Helmholtz decomposition $\mathbf{u} = \mathbf{u_p} + \mathbf{u_s}$, where $\mathbf{u_p}$ and $\mathbf{u_s}$ are respectively curl-free and divergence-free vector fields, two elastic waves can be retrieved: (1) a compression wave, which obeys to equation $\rho \frac{\partial^2 \mathbf{u_p}}{\partial t^2} = c_p^2 \Delta \mathbf{u_p}$, where $c_p = \sqrt{(K + \frac{4}{3}\mu)/\rho}$ is the compression wave speed; and (2) a shear wave, which obeys to equation $\rho \frac{\partial^2 \mathbf{u}_s}{\partial t^2} = c_s^2 \Delta \mathbf{u}_s$, where $c_s = \sqrt{\mu/\rho}$ is the shear wave speed. Hence, measuring the shear wave speed locally allows the estimation of the shear modulus (*i.e.*, elasticity).

## Shear wave speed estimation

Once optical images were acquired, displacements along X and Y were estimated using a 2 dimensional particle image velocimetry algorithm[6]. It is a Lucas-Kanade-based optical flow algorithm which assumes an affine (instead of a constant) displacement in each block. Shear wave speed was then estimated on the Y-displacement maps using a "passive" elastography algorithm[11, 12]. In this algorithm, the temporal cross-correlation between a point $(x_0, y_0)$ and all other points $(x, y)$ of the image is calculated to create a 2D+$t$ image $C_{(x_0,y_0)}(x, y; t)$ for each position considered $(x_0, y_0)$. Cross-correlation images typically look like a cross, showing a converging wave for $t < 0$, a refocusing at $t = 0$ with a maximum of amplitude, and a diverging wave for $t > 0$. Curvature of the focal spot is then evaluated on the resulting image, as the focal spot size is directly linked to the wavelength $\lambda$ of the shear wave. Shear wave speed $c_s$ is then estimated by multiplying the wavelength with the shear wave frequency $f$: $c_s = \lambda \times f$ - see supplementary materials for an illustration of the algorithm.

## SUPPLEMENTARY MATERIALS

## Validation of cytochalasin B experiment

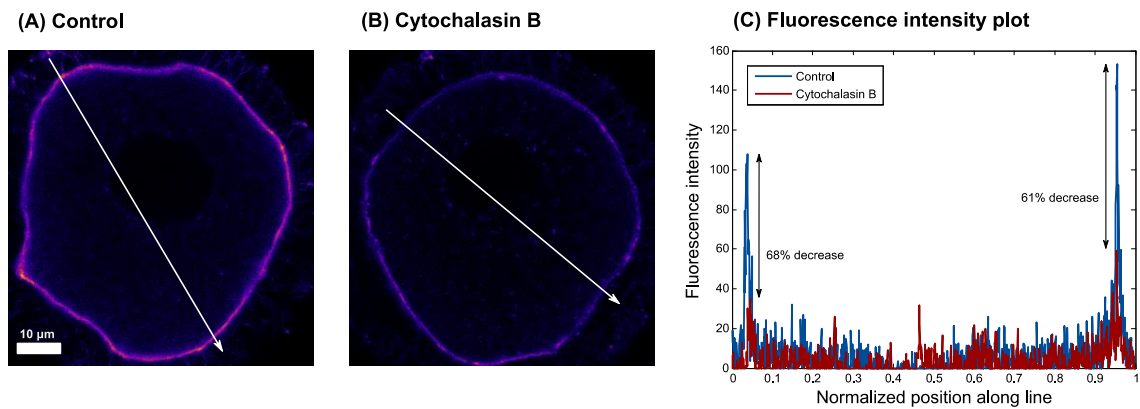

*Fig. 5: Demonstration that cytochalasin B disrupts the actin cortex in germinal-vesicle stage oocytes. Images show typical examples of control and cytochalasin-treated oocytes that were subsequently fixed and permeabilized, and the actin cortex labelled with alexa-phalloidin. The actin cortex, concentrated under the plasmalemma, is indicated as a pseudocolour image, warmer colours indicating greater alexa-phalloidin signal. Note that cytochalasin substantially diminishes, but does not completely remove the actin cortex. The graph on the right is a line intensity plot of fluorescence measured along the white lines, providing a quantitative readout of the loss of cortical actin.*

# Displacements along the horizontal direction

Similarly as the Y-displacement maps in Fig. 2, X-displacements over time are illustrated in Fig. 6 for the same excitation by the micropipette. We can see a good agreement between simulation and experiment, with comparable amplitudes and similar wavelength. The displacements are propagating mainly from the vibration pipette through the zona pellucida, because of the cell geometry. The experimental and simulated results have nevertheless some differences, especially at the vibration pipette location, mainly due to two reasons: (1) Experimental displacements calculated in the vicinity of the pipette are highly unreliable, whereas the simulation considered accurate displacements; and (2) X-displacements are about two times lower in amplitude than Y-ones, leading to a lower signal-to-noise ratio.

X-displacements maps were however not used to compute elasticity, as the amplitude was smaller, especially in the middle of the cell, so that it lead to rather unreliable results.

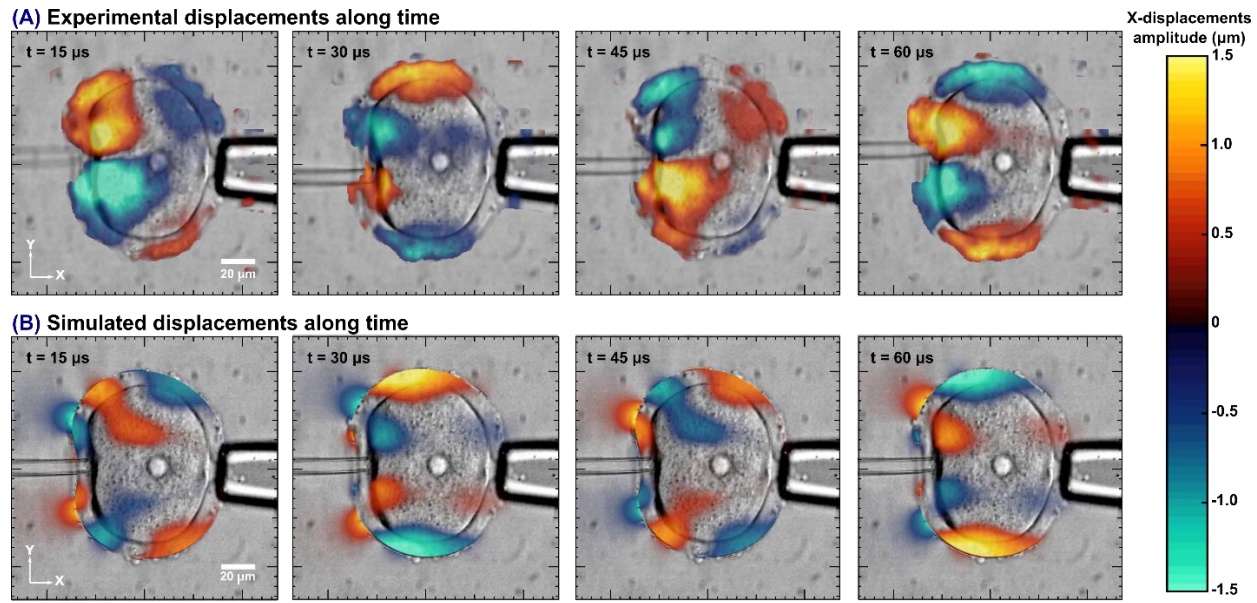

*Fig. 6: Experimental (A) and simulated (B) X-displacement maps, at t = 15, 30, 45 and 60 µs respectively, superimposed on the optical images of the cell. White bars on the left-bottom of each panel indicate a size of 20 µm. We can see displacements with an amplitude approximately from -1.5 to 1.5 µm, propagating from the vibrating pipette through the external layer of the cell.*

# Passive elastography algorithm principle

The "passive" elastography algorithm consists in a few steps, as illustrated in Fig. 7. It is based on: (1) estimating the displacement between each image over time, using the particle image velocimetry algorithm; (2) calculating the temporal cross-correlation between a point $(x_0, y_0)$ and all other points $(x, y)$ of the image; (3) measuring the focal spot size using the local curvature to obtain the local shear wavelength, so that by multiplying this wavelength with the shear wave frequency one can calculate the shear wave speed, and hence the shear modulus.

If the framerate is too low to track shear waves, the passive elastography algorithm will be able to produce an elasticity map but the results won't be quantitative[12].

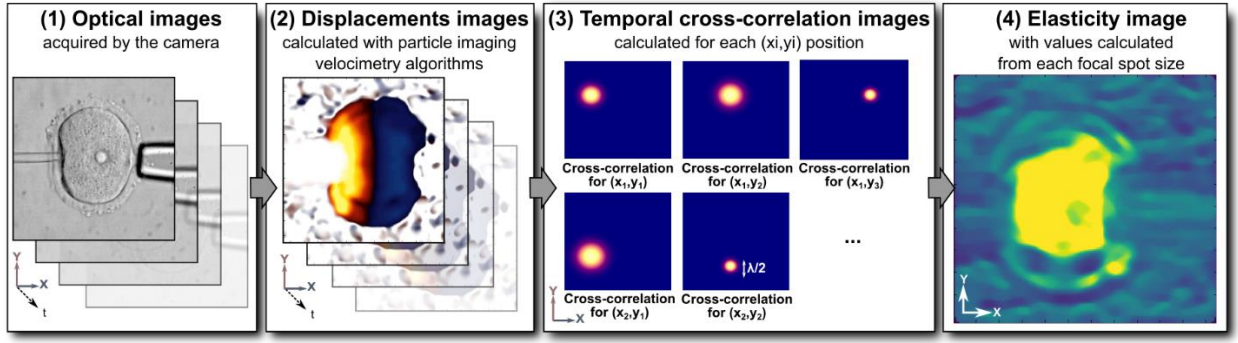

*Fig. 7: Illustration of the "passive" elastography algorithm applied on cells. The elasticity estimation consists in four steps: (1) acquisition of the optical images; (2) Estimation of the displacements between images; (3) Computation of the temporal cross-correlation of each image; and (4) estimation of the elasticity from the focal spot size.*

## Segmentation illustration

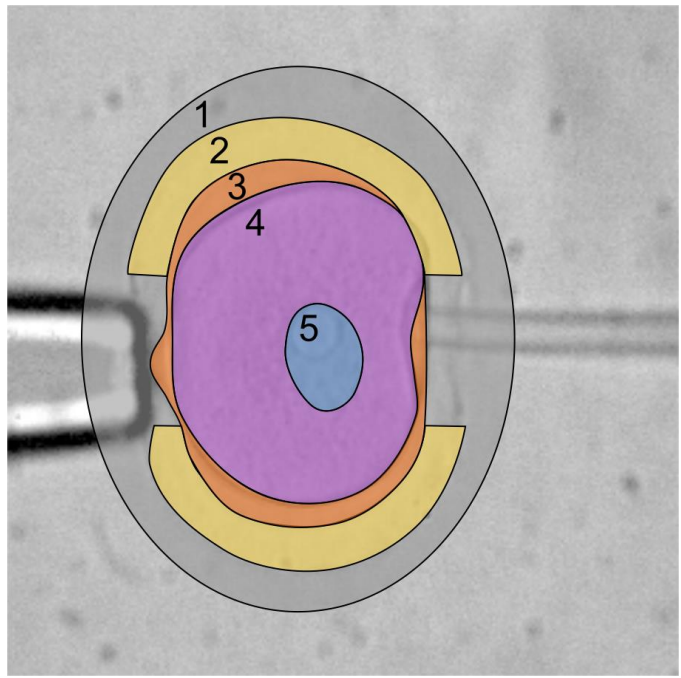

*Fig. 8: Illustration of the segmentation. 1: extracellular fluid; 2: zona pellucida; 3: perivitelline space; 4: cytoplasm; 5: nucleus; 2+3+4+5: whole cell.*

## Acknowledgments

Pol Grasland-Mongrain was recipient of a post-doctoral fellowship award from the Natural Sciences and Engineering Research Council of Canada (NSERC). Simon Bernard received a MEDITIS post-doctoral fellowship of NSERC provided by the Institute of Biomedical Engineering of the École Polytechnique and University of Montréal. This work was partially financed by the Fonds de Recherche du Québec – Nature et Technologies (grant #PR-174387), and the Canadian Institutes of Health Research (grants #MOP-84358 and #MOP-142334). The authors would like to thank Sylvain Bossé from Phantom Research for the formation on how to use the high-speed camera, and Julian Garcia for help on statistics.

## Author contributions

SC, GC and PGM designed the experiments. GC and GF provided financing and material support. PGM, SN and LGP conducted the experiments. SB and PGM conducted the simulations. PGM, SC and AZ analyzed the elasticity data. PGM produced the figures. PGM, GC, GF and SC wrote the manuscript. All authors discussed the results and reviewed the manuscript.